\documentclass[pra,twocolumn,superscriptaddress,showpacs,amsmath,amssymb]{revtex4-1}
\usepackage[dvipdfmx]{graphicx,color}
\usepackage{subfigure}
\usepackage{natbib}
\usepackage{bm}
\usepackage{siunitx}
\usepackage[normalem]{ulem}

\def\U#1{{\rm #1}} 
\def\u#1{_{\rm #1}}

\newcommand{\expect}[1]{\langle #1 \rangle}

\newcommand{\no}[1]{\left\langle: #1 :\right\rangle}  
\newcommand{\Gn}[2]{G^{(#1)}_{\rm #2}}
\newcommand{\Gnp}[2]{G^{\U{ph}(#1)}_{\rm #2}}
\newcommand{\gn}[2]{g^{(#1)}_{\rm #2}}
\def\tr{{\rm tr}}
\def\V2{{V_2^{(\U{ph})}}}
\def\X2{{X_2^{(\U{ph})}}}

\begin{document}
\title{
  Wave-particle duality of light appearing in the intensity interferometric situation
}
\author{Rikizo Ikuta}
\affiliation{Graduate School of Engineering Science, Osaka University,
Toyonaka, Osaka 560-8531, Japan}
\affiliation{
  Center for Quantum Information and Quantum Biology, 
  Osaka University, Osaka 560-8531, Japan}

\begin{abstract}
  We show a wave-particle duality of light and its complementary relation 
  in the context of the intensity interference 
  measured by intensity correlation measurement, 
  especially for the case of the second-order intensity interference 
  observed in the Hong-Ou-Mandel interferometer. 
  Different from the complementary relation
  appearing in the interference based on the phase coherence
  like in the Young's double-slit interferometer, 
  the complementary relation in the intensity interference 
  has a gap between classical and nonclassical lights. 
  This reveals a new nonclassical nature of light 
  where both wave and particle properties are classically understandable. 
  We further extend the wave-particle duality and the complementarity 
  to higher-order intensity interferometric situations. 
\end{abstract}
\maketitle

\section{Introduction}
Wave-particle duality of a quantum object is the heart of quantum mechanics~\cite{Feynman1965}.
Complementarity of the duality of a single photon~\cite{Zeilinger2005},
stemmed from Bohr's complementarity principle~\cite{Bohr1928}, 
has been quantitatively studied through double-slit experiments~\cite{Wootters1979, Greenberger1988, Mandel1991, Englert1996}.
The key idea of the analysis is to investigate a relation 
between visibility $V$ of the interference fringe as the wave-like property 
and so-called which-path information $D$ of the photon coming from the two arms 
as the particle-like property. 
There is a constraint between them known as 
the complementary relation as $D^2 + V^2\leq 1$. 
So far, the complementary nature of photons has been widely investigated 
from various viewpoints such as 
delayed-choice experiments~\cite{Jacques2007,Kaiser2012,Peruzzo2012} 
and interference experiments of partially incoherent light~\cite{Lahiri2017}
and composite systems~\cite{Jaeger1993,Jaeger1995,Herzog1995,Jakob2007}
as well as theoretical works related to the uncertainty principle~\cite{Scully1991,Wiseman1995,Coles2014}, quantum coherence~\cite{Bera2015} and entanglement~\cite{Qian2018}.
Recently, the concept of the wave-particle duality was expanded 
to higher-order interferometric situations~\cite{Huang2013,Huang2013-2}. 
In spite of having such a long history, to the best of our knowledge, 
all of the previous studies are related to phase correlation of optical states.
In the present day, 
the importance of intensity correlations is broadly known
in the sense of both interference and statistical distribution of photons~\cite{Glauber1963,Loudon2000}. 
In particular, for second-order intensity correlations, 
the intensity interference measured by the Hong-Ou-Mandel~(HOM) interferometer~\cite{Hong1987}
as the wave-like property 
and the autocorrelation function measured by the Hanbury-Brown and Twiss~(HBT) setup~\cite{Brown1956,Kimble1977}
as the particle-like property show the distinctive quantum features, 
and play integral roles in modern photonic quantum information technologies. 
However, these quantities were independently studied, 
and the wave-particle duality
in the intensity interferometric situation has never been discussed. 

In this study, we show the wave-particle duality of light
in the context of the intensity interferometric situation 
based on intensity correlation measurement, 
especially for the second-order intensity interference. 
By introducing the interference visibility for the intensity interference 
and which-path information in the situation, 
we derive the complementary principle 
between the visibility and the which-path information. 
The two complementary quantities are observed
by switching the measurement apparatus based on intensity correlation measurement. 
In the case of the second-order intensity interference, 
the quantities are measured by switching the HOM interferometer and the HBT setup
without changing the input setting. 
We show that the complementary principle of the duality is analogous 
to those of the dualities appearing in the phase-based interferometric situations 
in the classical wave theory, but can be violated in the quantum theory. 
This shows that compared with the classical light, 
the nonclassical light is allowed to 
have clear particle-like and wave-like properties simultaneously. 
In addition, the violation can be achieved 
even if the light has classically describable wave-like and particle-like properties. 
This fact reveals the new nonclassical interferometric nature of light 
which never be found from either wave-like or particle-like property alone. 

\section{Wave-particle duality in the interference by phase correlation}
\begin{figure}[t]
 \begin{center}
         \scalebox{1}{\includegraphics{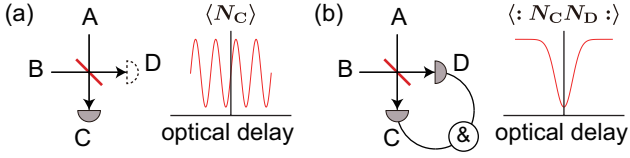}}
    \caption{
      Experimental setups and corresponding measurement results
      (a) for first-order interference 
      based on intensity measurement by a detector 
      such as the double-slit or the Mach–Zehnder interferometer,
      and
      (b) for second-order interference 
      based on intensity correlation measurement by two detectors 
      such as the HOM interferometer.
      The result in (b) is for an input light having no phase correlation. 
    }
    \label{fig:concept}
 \end{center}
\end{figure}

We review the complementary relation of the wave-particle duality 
in the interferometric situation based on the phase coherence. 
The first-order interference is represented by 
the double-slit experiment~\cite{Greenberger1988, Mandel1991, Englert1996}. 
The setup equivalent to the double-slit interferometer 
is shown in Fig.~\ref{fig:concept}~(a).
In the interferometer, two input lights coming from modes A and B
are mixed at a half beamsplitter~(HBS), 
and measured by detector $\U{D}\u{C}$ or $\U{D}\u{D}$. 
The detection probability $P\u{C(D)}$ at $\U{D}\u{C(D)}$ 
is proportional to the average photon number~(intensity) of light. 
By omitting the detection efficiency, 
$P\u{C}(=\expect{N\u{C}})$ is described by 
\begin{eqnarray}
  P\u{C} &=& \frac{1}{2}\left( \Gn{1}{AA} + \Gn{1}{BB} 
                      - 2|\Gnp{1}{AB}|\cos\theta_1\right),
\label{eq:fringe}
\end{eqnarray}
where $N_{i}:=a_i^\dagger a_i$ is the number operator of mode $i=$A, B, C 
with the annihilation operator $a_i$,
$\Gn{1}{AA(BB)}:=\expect{a\u{A(B)}^\dagger a\u{A(B)}}$ 
and 
$\Gnp{1}{AB}:=\expect{a\u{A}^\dagger a\u{B}}$ 
are the first-order correlation functions 
between modes $i$ and $j$~\cite{Glauber1963}, 
and $\theta_1 := \arg\Gn{1}{AB}$.
We used the transformation of the HBS as $a\u{C}\rightarrow (a\u{A}-a\u{B})/\sqrt{2}$. 

The which-path information $D_1$ is defined by 
magnitude of difference between the detection probabilities at $\U{D}\u{C}$ 
when only one half of light coming from the two input ports is entered,
which is $|\Gn{1}{AA} - \Gn{1}{BB}|/2$, with normalization by their total amount as
\begin{eqnarray}
  D_1 &:=& \frac{|\Gn{1}{AA} - \Gn{1}{BB}|}{\Gn{1}{AA} + \Gn{1}{BB}}. 
  \label{eq:D1}
\end{eqnarray}
On the other hand, the visibility $V_1^{(\U{ph})}$ is defined 
by the amplitude $|\Gnp{1}{AB}|$ of the interferometric fringe 
with the normalization factor used in Eq.~(\ref{eq:D1}) as 
\begin{eqnarray}
  V_1^{(\U{ph})} &:=& \frac{2|\Gnp{1}{AB}|}{\Gn{1}{AA} + \Gn{1}{BB}}. 
           \label{eq:V1}
\end{eqnarray}
From Eqs.~(\ref{eq:D1}) and (\ref{eq:V1}),
we obtain the complementary relation between $D_1$ and $V_1^{(\U{ph})}$~~\cite{Wootters1979, Greenberger1988, Mandel1991, Englert1996} as 
\begin{eqnarray}
  X_1^{(\U{ph})} &:=& D_1^2 + V_1^{(\U{ph})}{}^2 \\
      &=& 1 + 4\frac{|\Gnp{1}{AB}|^2 - \Gn{1}{AA}\Gn{1}{BB}}
          {\left( \Gn{1}{AA} + \Gn{1}{BB}\right)^2} \leq 1. 
  \label{eq:X1}
\end{eqnarray}
The last inequality is obtained as follows: 
For $O_n := \sum_{i=\U{A},\U{B}}\lambda_ia_i^n$ with complex values of $\lambda_i$,
$\tr(\rho O_n^\dagger O_n)
= \sum_{i=\U{A},\U{B}}\lambda_i^*\expect{ a_i^{\dagger n}a_j^n} \lambda_j \geq 0$
is satisfied. This means that the $2\times 2$ matrix $\{ \expect{ a_i^{\dagger n}a_j^n} \}$
is positive. 
As a result, due to the positivity of its determinant as
\begin{eqnarray}
  \Gn{n}{AA} \Gn{n}{BB} - |\Gnp{n}{AB}| \geq 0,
  \label{eq:consph}
\end{eqnarray}
Eq.~(\ref{eq:X1}) is derived for $n=1$,
where $\Gn{n}{AA(BB)}:= \expect{ a\u{A(B)}^{\dagger n}a\u{A(B)}^n}$
and $\Gnp{n}{AB}:= \expect{ a\u{A}^{\dagger n}a\u{B}^n}$. 

The complementarity was generalized to the $n$th-order interference~\cite{Huang2013},  
in which the which-path information $D_n$ and the fringe visibility
$V_n^{(\U{ph})}$ are defined by 
\begin{eqnarray}
  D_{n} &:=& \frac{| \Gn{n}{AA} - \Gn{n}{BB} |}{\Gn{n}{AA} + \Gn{n}{BB}},
  \label{eq:D2}\\
  V_n^{(\U{ph})} &:=& \frac{2|\Gnp{n}{AB}|}{\Gn{n}{AA} + \Gn{n}{BB}}. 
\label{eq:V2ph}
\end{eqnarray}
The quantities are obtained by $n$-fold coincidence photon detections 
without single-count measurements. 
From Eq.~(\ref{eq:consph}), the quantities satisfy the complementary relation as
\begin{eqnarray}
  X_{n}^{(\U{ph})} &:=& D_{n}^2 + V_n^{(\U{ph})}{}^2\\
  &=& 
      1 + 4\frac{|\Gnp{n}{AB}|^2 - \Gn{n}{AA}\Gn{n}{BB}}{\left( \Gn{n}{AA} + \Gn{n}{BB}\right)^2}
      \leq 1. 
\end{eqnarray}

\section{Wave-particle duality in the second-order intensity interference}
From here, we consider the intensity interferometric situation 
which is the main topic of this paper. 
Different from the double-slit experiment, 
the higher-order interference based on the intensity correlation measurement
includes the effects of not only the phase correlation but also the intensity correlation. 
We first consider the second-order interferometric experiment 
with the HOM interferometer as shown in Fig.~\ref{fig:concept}~(b). 
The setup is based on the intensity correlation~(coincidence) measurement 
by detectors $\U{D}\u{C}$ and $\U{D}\u{D}$. 
The coincidence probability measured by the interferometer 
is proportional to $\no{N\u{C}N\u{D}}$ with the use of the normal ordering, 
and it reflects the intensity correlation 
$\Gn{2}{AB}:=\expect{a\u{A}^{\dagger }a\u{B}^{\dagger } a\u{B}a\u{A}}=\no{N\u{A}N\u{B}}$
as well as the phase correlation $\Gnp{2}{AB}$. 
In fact, the coincidence probability $P_\parallel$ 
when modes of the photons coming from modes A and B are perfectly indistinguishable after the HBS 
is described by 
\begin{eqnarray}
  P_\parallel &:=& P_\perp - \frac{1}{2}\left(\Gn{2}{AB} + |\Gnp{2}{AB}|\cos\theta_2\right)
    \label{eq:P01}\\
      &=&\frac{1}{4}\left(\Gn{2}{AA}+\Gn{2}{BB} - 2 |\Gnp{2}{AB}|\cos\theta_2 \right),
  \label{eq:P0}
\end{eqnarray}
where $P_\perp := \no{(N\u{A}+N\u{B})^2}/4$ is the probability 
when modes of the photons coming from A and B are perfectly distinguishable after the HBS 
as the classical objects. 
In experiments, the phase correlation $|\Gnp{2}{AB}|$ is observed 
as the oscillation of the coincidence probability~\cite{Ou1988,Shih1994,Dowling2008}. 
On the other hand, the intensity correlation $\Gn{2}{AB}$ is observed 
by the depth of $P_\parallel$ from $P_\perp$ with excluding the effect of 
the phase correlation~\cite{Rarity2005,Riedmatten2003,Ikuta2013-2,Tsujimoto2017},
which is known as the HOM dip as shown in Fig.~\ref{fig:concept}~(b). 

In the situation, we consider the which-path information $D_2$
in Eq.~(\ref{eq:D2}) as the particle-like property. 
It is measured by blocking one of the input ports corresponding to the HBT setup. 
$D_2$ is understood as the probability of guessing which way the light came from, 
by the detection statistics 
based on the coincidence measurement
without mixing the light coming from the two input ports.
This physical meaning is the same as $D_1$ except for the difference
between the single-count and coincidence measurements. 

We note that the which-path information is sometimes called {\it distinguishability}. 
But the word is confusing because the distinguishability is conventionally used
for characterizing the mode-matching degree of the photons after {\it mixing} at the HBS. 
The concept is rather related to the intensity interference effect in Eq.~(\ref{eq:P01}) 
and the visibility we introduce below. 

We define the interference visibility $V_2$ 
for the intensity correlation by depth $\Gn{2}{AB}/2$ of the HOM dip 
with the normalization as 
\begin{eqnarray}
  V_2 := \frac{2\Gn{2}{AB}}{\Gn{2}{AA} + \Gn{2}{BB}}. 
  \label{eq:V2I}
\end{eqnarray}
While the definition is different from 
the commonly used HOM visibility as $V\u{HOM}:= 1 - P_\parallel/P_\perp$~\cite{Riedmatten2003,Ikuta2013-2,Tsujimoto2017}, 
they are equivalent because $V\u{HOM} = (1 + V_2^{-1})^{-1}$ is satisfied. 

From Eqs.~(\ref{eq:D2}) and (\ref{eq:V2I}), 
we see that the quantities $D_2$ and $V_2$ satisfy the complementary relation as 
\begin{eqnarray}
  X_2 &:=& D_2^2 + V_2^2\\
    &=& 1 +
        4\frac{(\Gn{2}{AB})^2 - \Gn{2}{AA}\Gn{2}{BB}}
        {\left( \Gn{2}{AA} + \Gn{2}{BB}\right)^2} 
        \label{eq:X2}\\
  &\leq & 1
\end{eqnarray}
for 
\begin{eqnarray}
  (\Gn{2}{AB})^2\leq \Gn{2}{AA} \Gn{2}{BB}.
  \label{eq:cons}
\end{eqnarray}
The inequality is derived by the Cauchy–Schwarz inequality, 
but it makes well {\it only in the classical wave theory}. 
Different from the complementarity of the duality 
appearing in the phase interference as $X_{n}^{(\U{ph})} \leq 1$ 
for all of the light, $X_2 > 1$ is possible in the quantum theory. 
This means that compared with any light satisfying Eq.~(\ref{eq:cons}), 
the nonclassical light is allowed to 
have strong particle-like and wave-like signatures simultaneously 
in the context of the second-order intensity interference. 

To see the striking feature of the nonclassical complementarity $X_2 > 1$ in more detail, 
it is convenient to introduce the normalized second-order cross correlation function 
$\gn{2}{AB}:=\Gn{2}{AB}/(\Gn{1}{AA}\Gn{1}{BB})$ 
and the normalized autocorrelation function 
$g^{(2)}_{ii}:=G^{(2)}_{ii}/G^{(1)}_{ii}{}^2$
which characterizes the photon statistics of light in mode $i=$A, B 
measured by the HBT setup~\cite{Brown1956}. 
With the use of the quantities, we consider two typical examples as follows. 

For the first example, we assume that the lights at modes A and B 
are statistically independent as $\gn{2}{AB}=1$. 
In addition, we assume $\gn{2}{AA}$ and $\gn{2}{BB}$ take the same value as $\gn{2}{auto}$. 
In this case, $D_2$ is independent of $\gn{2}{auto}$ as 
$D_2 = |\zeta-\zeta^{-1}|/(\zeta + \zeta^{-1})$, where
$\zeta := \Gn{1}{AA}/\Gn{1}{BB} = \expect{N_A}/\expect{N_B}$ is the intensity ratio. 
On the other hand, the visibility depends on $\gn{2}{auto}$ 
as $V_2 =2(\zeta+\zeta^{-1})^{-1}\gn{2}{auto}{}^{-1}$. 
As a result, without changing the value of $D_2$, 
it is possible to obtain $V_2 \rightarrow \infty$, 
which leads to $X_2 \rightarrow \infty$, for $\gn{2}{auto}\rightarrow 0$. 
This means the nonclassical lights with $\gn{2}{auto} < 1$ 
show the larger interference signature 
than that of any classical light
even though they have the same value of which-path information $D_2$. 
We notice that this example includes the case 
where the input is two single photons with one in each input port. 
While $D_2$ for this state is not directly calculated from Eq.~(\ref{eq:D2}), 
it is determined by the asymptotic value for $\gn{2}{auto}\rightarrow 0$
as described above. 

\begin{figure}
 \begin{center}
         \scalebox{1.}{\includegraphics{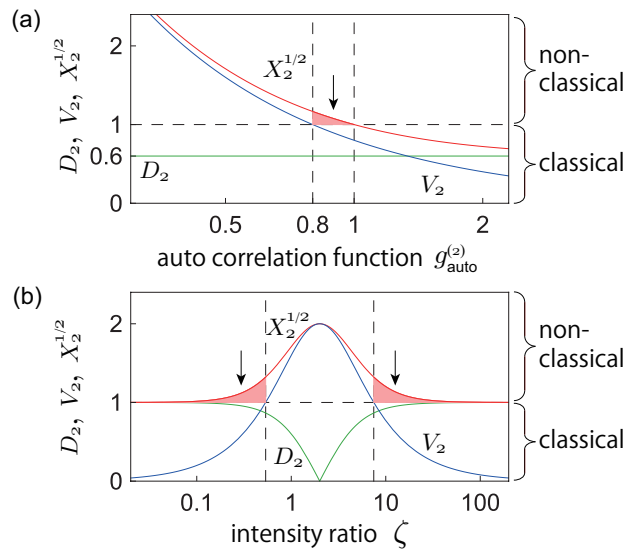}}
    \caption{
      Log–linear plots of $D_2$~(green), $V_2$~(blue) and $X_2{}^{1/2}$~(red). 
      (a) $\gn{2}{auto}$ dependencies 
      with assumptions of $\gn{2}{AB}=1$, $\gn{2}{AA}=\gn{2}{BB}=\gn{2}{auto}$ 
      and $\zeta=2$. 
      (b) $\zeta$ dependencies 
      with assumptions of $\gn{2}{AB}=\gn{2}{BB}=1$ and $\gn{2}{AA}=0.25$. 
      The areas marked with arrows show 
      the nonclassical complementarity $X_2 > 1$~($X_2{}^{1/2} > 1$)
      whereas the classical explained values of $D_2\leq 1$ and $V_2 \leq 1$.
    }
 \label{fig:example}
 \end{center}
\end{figure}
From the above example, 
one may guess that the nonclassical complementarity $X_2 > 1$ always comes from 
the nonclassical visibility $V_2 > 1$.
This is because 
$V_2$ is guaranteed to be $0\leq V_2 \leq 1$ only in the classical wave theory 
but takes $0\leq V_2 \leq \infty$ in the quantum theory
from $V_2^{-2}\geq \Gn{2}{AA}\Gn{2}{BB}/(\Gn{2}{AB})^2$, 
while $0\leq D_2 \leq 1$ is satisfied in the both theories. 
However, the second example considered below shows the counter-example of such an intuition. 

For the second example, 
we assume $\gn{2}{AB}=1$ and $\gn{2}{auto} = \gn{2}{AA} = \gn{2}{BB}$
similar to the first example. 
In addition, we assume $\gn{2}{auto} = 0.8$ and $\zeta = 2$. 
In this case, we see the classically explained which-path information
and the interference visibility as $D_2 = 0.6$ and $V_2 = 1$, 
the latter of which corresponds to the classical limit 
of the HOM visibility as $V\u{HOM}=0.5$. 
Nonetheless, the nonclassical complementarity appears as $X_2 = 1.36$. 
This nonclassicality is obtained only after considering $D_2$ as well as $V_2$.
The visualization of the relation among $D_2$, $V_2$
and $X_2$ is shown in Fig.~\ref{fig:example}~(a)
which includes the first example explained in the previous paragraph. 
This nontrivial situation appears in a more feasible experimental setting 
where a heralded single photon produced by spontaneous parametric down conversion 
and independently prepared laser light~($\gn{2}{AB} = \gn{2}{BB} =1$) are mixed 
like in Refs.~\cite{Rarity2005,Ikuta2013-2}. 
We show $\zeta$ dependency of the complementarity in Fig.~\ref{fig:example}~(b)
with assuming $\gn{2}{AA} = 0.25$. 

We remark that
the condition in Eq.~(\ref{eq:cons}) needed for the classical complementary principle $X_2 \leq 1$ 
was derived as a result of considering the mixing of the bosonic photons 
and newly introducing the second-order which-path information and the visibility. 
Interestingly,
the same inequality happens to appear 
as a condition for the requirement of 
the classically explained cross correlation function 
of separated photon pairs~\cite{Clauser1974,Walls2008}, 
despite the completely different physical situation from that considered in this paper. 

\section{Duality in the higher-order intensity interferometric situations}
We generalize the wave-particle duality of light to 
higher-order intensity interferometric situations 
based on $n$-fold coincidence detection after a BS 
which distributes input light from modes A and B to $n$ output modes equally. 
The interference effect includes $n-1$ types of the intensity correlations 
$G^{(n)}_{k,{\rm AB}}:=\no{ N\u{A}^kN\u{B}^{n-k} }$ for $k=1,\ldots, n-1$
and the phase correlation $\Gnp{n}{AB}$. 
For the interference related to $G^{(n)}_{k,{\rm AB}}$, 
We define the which-path information as 
\begin{eqnarray}
D_{n,k} := \frac{| \Gn{2k}{AA} - \Gn{2(n-k)}{BB} |}{\Gn{2k}{AA} + \Gn{2(n-k)}{BB}},
\end{eqnarray}
and the visibility as 
\begin{eqnarray}
  V_{n,k} := \frac{2G^{(n)}_{k,{\rm AB}}}{\Gn{2k}{AA} + \Gn{2(n-k)}{BB}}.
\end{eqnarray}
From the definitions, we obtain 
\begin{eqnarray}
  X_{n,k} &:=& D_{n,k}^2 + V_{n,k}^2\\
          &=& 1 +
        4\frac{G^{(n)}_{k,{\rm AB}}{}^2 - \Gn{2k}{AA}\Gn{2(n-k)}{BB}}{\left( \Gn{2k}{AA} + \Gn{2(n-k)}{BB}\right)^2}. 
              \label{eq:Xn} 
\end{eqnarray}
Similar to the complementary relation
$X_{2,1}(=X_2) \leq 1$ of the second-order interference for classical light, 
$X_{n,k}\leq 1$ for $n\geq 3$ is always satisfied in the classical wave theory 
due to the Cauchy–Schwarz inequality, but is not in the quantum theory.
In fact, $X_{n,k} > 1$ is achieved in the similar situations we gave above
as examples for $n=2$. 

From an experimental point of view, 
the interferometric setup to estimate the quantity $X_{n,k}$ 
will be more complicated for larger value of $n$.
One of the reasons is extraction of the interference effect caused by $V_{n,k}$
for desired $k$ would be difficult because the $n$-fold coincidence probability generally 
includes all of the $n$-th order interference effects. 
Another problem is the estimation of $D_{n,k}$. 
For the symmetric case of $n=2m$ and $k=m$, 
$D_{n,k}=D_n$ is estimated by $n$-fold coincidence measurement.
But $D_{n,k}$ does not correspond to $D_n$ in general. 
For $k \neq n/2$, 
$n$-fold coincidence measurement is not enough to estimate $D_{n,k}$. 
One of the solution to overcome this difficulty
is to replace a photon detector after each of $n$ output ports of the BS for mixing modes A and B 
by a HBS followed by two detectors. 
This setup can detect up to $2n$-fold coincidence events
without affecting the interferometric effect between input lights from modes A and B. 
As a result, both $D_{n,k}$ and $V_{n,k}$, resulting in $X_{n,k}$, could be estimated. 

\section{Conclusion}
In conclusion, we introduced the concept of the wave-particle duality of light 
in higher-order intensity interferometric situations especially 
for the second-order interference measured by the HOM interferometer. 
We showed that the complementary relation of the duality 
is analogous to that in the phase-based interference, 
but only in the classical wave theory. 
In the quantum theory, the relation is violated. 
The nonclassical complementarity highlights 
the clear wave-like behavior of nonclassical light 
compared with classical light having the same particle-like property, 
and in addition, reveals the nonclassical aspect of light 
which possesses classically understandable wave and particle properties. 

\section*{Acknowledgement}
The author thanks Y.~Tsujimoto, T.~Yamazaki, and T.~Kobayashi for giving me helpful comments. 
This work was supported by CREST, JST JPMJCR1671;
MEXT/JSPS KAKENHI Grants No. JP21H04445 and JP20H0183.

\end{document}